 \documentstyle[preprint,aps,prb]{revtex}

\def\bfp{{\bf p}}

\def\pl{\partial}

\begin{document}
\title{Ideal Bose gas in fractal dimensions and superfluid $^4$He
 in porous media}
\author{Sang-Hoon Kim,$^1$ Chul Koo Kim,$^2$ and Kyun Nahm$^3$}
\address{$^1$Division of Liberal Arts, Mokpo National Maritime University,
 Mokpo 530-729, Korea}
\address{$^2$Department of Physics and Institute for Mathematical Sciences,
 Yonsei University, Seoul 120-749, Korea}
\address{$^3$Department of Physics, Yonsei University, Wonju 220-710, Korea}
\date{\today}
\maketitle
\draft
\begin{abstract}
Physical properties of ideal Bose gas with the fractal
 dimensionality between $D=2$ and $D=3$ are theoretically investigated.  
Calculation shows that the characteristic features of the specific heat
 and the superfluid density of ideal Bose gas in fractal dimensions 
are strikingly similar to those of superfluid Helium-4 in porous media.
This result indicates that the geometrical factor is  dominant
over  mutual interactions in determining  physical properties of 
Helium-4 in porous media.
\end{abstract} 
\vspace{0.25in}
\newpage
\section{introduction}

Superfluid Helium-4 is a typical boson fluid to occur in nature and
 Bose condensation is believed to be the fundamental reason for such a
phenomenon.  Recently, superfluidity of liquid helium-4 in highly
connected porous structures (vycor glass, glass plate, xerogel, aerogel, 
graphite, fine powders, steel, German silver, plastic films, {\em etc}.)
has been observed and studied intensively. 
\cite{Fred,Brew,Bret,Fino,Steele,Rep1,Rep2,Rep3}
However, so far no satisfactory explanation
of experimental observations on these materials has been achieved.
\cite{Bowl,Khur,Gob,KT,Rep4,phil,Gasp,Will,Saar}
Since porous media can be treated as solids with fractal or non-integer
 dimensions, \cite{Gasp,Will,Saar,PfOb,PfWu,PfCo,Pf} we believe that 
it is imperative to study the extent of the dimensionality
 contribution to the superfluidity in order to understand the
experimental results.

In this paper, we examine the physical properties of ideal Bose gas
 with fractal dimensions between $D=2$ (thin film limit)
 and $D=3$ (the bulk limit).  The results will be 
compared with experimental results obtained from liquid Helium-4
in porous media.
Surprisingly, we find that  most of salient features of experimental
observations on liquid helium-4 in porous solids can be explained using
 the theoretical results from the ideal gas model in fractal dimensions.
This indicates that the  dimensionality contribution, 
not the mutual interactions, is the dominant factor for superfluidity
 in porous media.

\section{Ideal Bose gas in non-integer dimensions}
  The  ideal Bose gas system  at  integer dimensions was studied
 long time ago.\cite{Ziff}  However, study at fractal dimensions was 
carried out only quite recently.\cite{Pf,kim}
The  density of states which is essential to calculate 
 thermodynamic properties of the Bose gas in $D$-dimension,
 where $D$ is any real number, is given by \cite{PfOb}
\begin{equation}
\rho_D(E) = a_D E^{\frac{D}{2} - 1},
\label{14}
\end{equation}
where $a_D$ is the $D$-dimensional coefficient which is known as
\begin{equation}
a_D = \frac{V(D)}{\Gamma\left( \frac{D}{2} \right)}
  \left( \frac{m}{2 \pi \hbar^2} \right)^{\frac{D}{2}}.
\label{dos}
\end{equation}
Here, $\Gamma$ is the Gamma function,  $m$  the mass, and 
$V(D)$  the $D$-dimensional {\it measure} or volume.

In order to obtain physical quantities of the $D$-dimensional Bose gas,
it is necessary to obtain the grand partition function in $D$-dimension,
\begin{eqnarray}
\ln Q(z,v,t) &=& - \ln (1-z) - \int_0^\infty \ln (1-z e^{-\beta E})\rho_D (E) dE
\nonumber \\
&=&  - \ln (1-z) + \frac{V}{\lambda^D} g_{D/2 +1}(z)
\label{23}
\end{eqnarray}
where  $z$ is the thermodynamic fugacity defined by
  $z = e^{\beta \mu}$,  $E(\bfp) = p^2/2m$,
 and $\lambda(T)$ is the thermal wavelength defined 
by $\sqrt{2 \pi \hbar^2/mk_{B}T}$.
With the coefficient $a_D$, we can readily calculate the grand partition
 function in $D$-dimension.
The  $g_{s}(z)$ is the  Bose gas function defined by
\begin{equation}
g_s(z) = \sum_{n=1}^{\infty} \frac{z^n}{n^s}.  \\
\label{gsz}
\end{equation}
The coefficients $s$ and $z$ are  restricted to
$ s > 0 $ and $ 0 \leq z \leq 1 $ regions.
The Bose gas function  can be also extended to non-integer
dimensions,\cite{Robi} and has an integral expression,
\begin{equation}
g_s(z)  = - \frac {1}{\Gamma(s-1)} \int_0^\infty dx \, x^{s-2}
           \, \ln (1-z \, e^{-x}).
\label{git}
\end{equation}
Note that this  representation is valid only when $s > 1$.

Using the above expressions, we obtain the average number of particles
 in $D$-dimension, 
\begin{eqnarray}
N &=& z \frac{\partial}{\partial z} \ln Q(z,v,t)
\nonumber \\
& = & \sum_{\bf p \neq \bf 0} <n_{\bf p}> + <n_{\bf 0}>
\nonumber \\
  & = & \frac {V}{\lambda^{D}} g_{D/2}(z) + \frac {z}{1-z}.
\label{anp}
\end{eqnarray}
We use the above  equations to calculate thermodynamic properties
in fractal dimensions in the following.

\section{Physical properties of ideal Bose gas in fractal dimensions
and application to liquid Helium-4 in porous media}
 Before we go into detailed calculations, we briefly summarize
experimental findings on superfluid Helium-4 in porous media,
 since we are interested in whether the dimensionality
 considered above plays any role in real situations. 

 It is known  that physical properties of superfluid Helium-4
 in porous media show quite puzzling behaviors.
\cite{Rep1,Rep2,Rep3,Rep4,phil}
Upon reducing the lab variables of porous media such as ``pore filling,"
``thickness," ``number of layers," or ``coverage,"
 the capacity curves show the following generic behaviors independent
of the porous media or substrates:
(i) The critical temperature for superfluidity onset shifts downwards
almost linearly.
(ii) The bulk-like sharp cusp disappears and the peak gets smaller.
  Furthermore, there appears a systematic crossing at low temperatures.
(iii) The shape of the curve gets rounder and, eventually, becomes flat.
These  experimental findings are summarized in FIG. 1.  These figures are 
reproduced from references [1-4], so that comparison with the theory can be 
made transparent.
Here, we observe that the above experimental findings appear independent of 
detailed nature of porous media and also of interactions between particles.  
This fact strongly suggests that, at least, the qualitative nature of the 
above behavior may originate from geometric factors.
(iv) Another interesting physical property is the superfluid density shown
 in FIG. 2.  The sample E has smaller density, which implies more
 connectivity of porous structure than the sample F in FIG. 2(a).
  It is shown that the curves from different media do not
cross one another.  This again indicates that the superfluid density
is strongly dependent on the porous structure.

Numerous  theories have been suggested to explain the above experimental
 observations. \cite{Bowl,Khur,Gob,KT,Gasp,Will,Saar} 
 However, so far no completely successful theory has emerged.
For example, the KT model of the vortex mediated transition in
 two-dimensional space  gives an excellent explanation of  question (i)
for very thin films of one or two layers but not for other samples.

We now calculate the physical properties in fractal dimensions 
and compare the results with the experimental findings summarized above.
Fractal dimensionality measures disorderness in terms of the connectivity
 of the system.

(i) The critical temperature, $T_c$, for the superfluidity of the ideal Bose
 gas in fractal space can readily obtained from Eq. (\ref{anp}) to be
 given by \cite{Rep4,kim}
\begin{equation}
k_{B}T_{c} = \frac{2 \pi \hbar^2}{m}
\frac {1}{[ vg_{D/2}(1) ]^{\frac{2}{D}}}.
\label{ctem}
\end{equation}
The almost linear behavior of  $T_c$ as a function of dimensionality is 
plotted in FIG. 3.
Necessary  parameters are taken from reference 25.
The formula can be simplified  when  $D$ approaches to 2 to be given by
\cite{Grad} 
\begin{equation}
 T_c  \sim  \left| \frac{D}{2}-1 \right|.
\label{Tc}
\end{equation}
 The results are surprisingly in good agreement with the experimental findings
 shown in FIG. 1.  Even though  we completely neglected the interactions
between  Bose particles, the $D \rightarrow 2$ limit has the same form
 with the KT theory prediction.\cite{KT}  
  Note that  the KT theory cannot account the experimental results  for
 thick films, but the present  $D \rightarrow 3$ limit agrees with the
correct bulk value.

(ii) The specific heat of the ideal Bose gas in fractal space can readily
 obtained, too, from the grand partition function $Q$ in Eq. (\ref{23}).
(a) when $T \leq T_c$
\begin{equation}
\frac{C_V(T)}{Nk_B} = \frac{D}{2} \left( \frac{D}{2}+1 \right)
\frac{v}{\lambda^D} g_{D/2+1}(1),
\label{cvL}
\end{equation}
and, (b) when $T > T_c$
\begin{equation}
\frac{C_V(T)}{Nk_B} = \frac{D}{2} \left( \frac{D}{2}+1 \right)
\frac{v}{\lambda^D}
g_{D/2+1}(z) - \left( \frac{D}{2} \right)^2 \frac{g_{D/2}(z)}{g_{D/2-1}(z)}.
\label{cvH}
\end{equation}
The $C_V(T)$ curves for several values of fractal dimensions 
are plotted as functions of temperatures in FIG. 4. 
 We observe that the cusp disappears when the dimension is less than 3, 
and the peak height becomes smaller with decreasing dimensionality.  
Also, FIG. 4 shows that there exists a systematic crossover at low
 temperature regions.  These results are in excellent agreement with
 the experiments shown in FIG. 1.

(iii) In FIG. 4, we observe that the peaks of the specific heat curves
 become less and less prominent and, eventually, the curves become flat
 with decreasing  dimensionality.  We show that this behavior originates
 from a hidden hierarchy in the superfluidity transition with fractal
 dimensions.  The first temperature derivative of the specific heat can be
 readily obtained from  Eqs. (\ref{cvL}) and (\ref{cvH}).
(a) when $T \leq T_c$
\begin{equation}
\left( \frac{\partial}{\partial T} \right)_V\frac{C_V(T)}{Nk_B}
= \left(\frac{D}{2}\right)^2 \left( \frac{D}{2}+1 \right)
\frac{v}{\lambda^D} \frac{g_{D/2+1}(1)}{T},
\label{ccvL}
\end{equation}
(b) when $T > T_c$
\begin{eqnarray}
\left( \frac{\partial}{\partial T} \right)_V \frac{C_V(T)}{Nk_B}
&=& -\left( \frac{D}{2}\right)^2 \frac{1}{T}
\frac{g_{D/2}(z)}{g_{D/2-1}(z)}
\nonumber \\
&\times& \left\{ 1- \left( \frac{D}{2}+1 \right)
\frac{g_{D/2+1}(z) g_{D/2-1}(z)}{[g_{D/2}(z)]^2}
+ \frac{D}{2}\frac{g_{D/2}(z) g_{D/2-2}(z)}{[g_{D/2-1}(z)]^2} \right\}.
\label{ccvH}
\end{eqnarray}
The first derivative of $C_V$ is plotted in FIG. 5 in arbitrary unit. 
The curves again show the shift of $T_c$ with decreasing  dimensionality.
  Also, it is shown that the discontinuity of
 $(\partial C_V(T) /\partial T)_V $  disappears with decreasing 
 dimensionality.  In order to investigate this behavior more closely, 
we studied  the relation between the continuity of higher derivatives of $C_V$
 and the fractional dimensionality.
Taking higher derivatives on $C_V$ and considering the behavior at $T_c$, 
we obtain the following relation:
\begin{equation}
 \lim_{T \rightarrow T_c}
  \left[   \left( \frac{\pl}{\pl T} \right) ^n_V C_V^-(T)
   -  \left( \frac{\pl}{\pl T} \right) ^n_V C_V^+(T) \right]
= \lim_{\eta \rightarrow 0}
   \sum_{j=1}^n a_{nj} \, \eta ^{j+2-\frac{(j+1)D}{2}},
 \label{crt}
\end{equation}
where the coefficients $a_{nj}$ are finite constants, and
  $C_V^-$ and $C_V^+$ are the specific heats  below  and above $T_c$.
This formula can also be proved by mathematical induction method 
(see APPENDIX).
The hierarchy of the superfluidity transition with the fractal
 dimensionality obtained from the above formula is summarized at  TABLE 1.
This table explains the physical origin of the roundness tendency of the specific heat curves with decreasing the dimensionality.

(iv) In real system the superfluid fraction is different from the 
condensate fraction, but the structure of the condensate fraction may
shed some clue for the superfluid fraction.  The condensate fraction
in $D$-dimension can  readily be obtained from
Eqs. (\ref{anp}) and (\ref{ctem})  to be given by
\begin{equation}
\frac{n_0}{n} = 1 - \left[ \frac{T}{T_c(D)}\right]^{\frac{D}{2}}
\label{41}
\end{equation}
The curves are shown for several fractal dimensions in FIG. 6.
It is shown that the curves with  higher dimensions have  higher $T_c$
 and the curves do not cross each other.
The  FIG. 6 gives the information that the sample E has a higher fractal
 dimension than the sample F in FIG. 2(a).
Also,  the aerogel has the highest fractal dimension,
 the xerogel is the next, and the vycor has the lowest one in FIG. 2(b).
 Of course, they are all less than the bulk value of $D = 3$.
The qualitative features of the theoretical curves are in excellent
 agreement with the experimental results in FIG. 2.
Even the tail structure at $\rho_0 \rightarrow 0$  is reproduced.

\section{discussions}
Physical properties of the ideal Bose gas in  fractal dimensions 
are studied theoretically.  The results are compared with the experimental 
findings from the superfluid  Helium-4 in porous media. 
 It is found that the main characteristics of the experimental results 
are in excellent agreement with the theoretical results obtained  from 
the simple ideal Bose gas in fractal dimensions.  
Such a good agreement may not be totally unexpected.
Careful examination of the experimental results reveals that the salient
 features  are independent of materials and mutual interactions, thus
 suggesting that the dominant contributions should come from 
 geometrical factors.
  Another important point of the theoretical study is the existence
 of a hierarchy of the superfluidity transitions with changing fractal
 dimensionality.  The peak of heat capacity is not necessary for the 
superfluid transition in porous media.
It will be  interesting to study how the such physical properties
 modified when interactions are included in the calculations.

 \acknowledgments
The author thanks to  P. Pfeifer and C.E. Campbell for useful discussions. 
This research was partially supported by the Mokpo National Maritime
 University and the Korean Research Foundation (98-015-D00061).
\begin{appendix}
\section{}

Here, we prove Eq. (\ref{crt}) using the mathematical induction method. 
 First, we give some  useful relations of the Bose gas function
which are needed for the derivation:
\begin{eqnarray}
g_s(z)   \sim &        z
       & \hspace{.4in}    : z \rightarrow 0^+,
\nonumber  \\
         \sim & \Gamma (1-s)(-\ln z)^{s-1} + \zeta (s)
       & \hspace{.4in}    : z \rightarrow 1^-,
\label{sim2}
\end{eqnarray}
where $\zeta$ is the Riemann-zeta function.
\begin{equation}
\left( \frac {\pl z}{\pl T} \right)_V =
- \frac{D}{2}\,\frac {z}{T} \frac {g_{D/2}(z)}{g_{D/2-1}(z)}.
\label{210}
\end{equation}
  We can use above relations  for $(\pl \eta/ \pl T)_V$, too.
We put  $z = e^{-\eta}$, then $z \rightarrow 1$  as $\eta \rightarrow 0$. 

We introduce a differential operator defined by
\begin{equation}
\Delta ^n(T) \equiv \left(\frac {\pl }{\pl T}\right)^n_V C_V^-(T) -
                     \left(\frac {\pl }{\pl T}\right)^n_V C_V^+(T)
\label{del}
\end{equation}
where $C_V^-$ and  $C_V^+$ are the specific heat below and above $T_c$.
For convenience, we drop the limit notation of
`$\lim_{\eta \rightarrow 0}$ (or $T \rightarrow T_c$)' during the proof.

(i) When $n=1$,
\begin{equation}
\Delta^1(T_c) =  a_{11}  \, \eta ^{3-D}.
\end{equation}
This is clearly true from Eq. (\ref{crt}) and the known result for $D=3$.

(ii) We assume  that Eq. (\ref{crt}) is true for any positive integer $k$.
  Then
\begin{equation}
\Delta^k(T_c)=  \sum _{i=1}^k a_{ki} \, \eta ^{i+2-\frac {i+1}{2}D}.
\end{equation}
Using Eqs. (\ref{sim2}) and  (\ref{210}), we obtain
\begin{eqnarray}
\Delta ^{k+1}(T_c)
&=& \sum _{i=1}^k a_{ki} \, (i+2 - \frac {i+1}{2}D) \,
\eta ^{i+1-\frac {i+1}{2}D} \, \left( \frac{\partial \eta}{\pl T} \right)_V
\nonumber \\
&=& \sum _{i=1}^k a_{ki} \, (i+2 - \frac {i+1}{2}D) \,
\frac{D \, \zeta(\frac{D}{2})}{2 \, T_c \, \Gamma(2- \frac{D}{2})} \,
\eta ^{i+3-\frac {i+2}{2}D}
\nonumber \\
&=& \sum _{j=2}^{k+1} a_{k+1,j-1} \, (j+1 - \frac {j}{2}D) \,
\frac{D \, \zeta(\frac{D}{2})}{2 \, T_c \, \Gamma(2- \frac{D}{2})} \,
\eta ^{j+2-\frac {j+1}{2}D}
\nonumber \\
&=& \sum _{j=2}^{k+1} a_{k+1,j} \, \eta ^{j+2-\frac {j+1}{2}D}.
\label{dkt}
\end{eqnarray}
The $a_{ki}$ satisfies the recurrence relation
\begin{equation}
a_{k+1,j}=
\frac { (j+1 - \frac {j}{2}D) D \, \zeta(\frac{D}{2})}
{2 \, T_c \, \Gamma(2- \frac{D}{2})}  \, a_{k+1,j-1}
\label{arl}
\end{equation}
where  $j = 2,3,4,...,k+1.$

Therefore,  (i) and (ii) enables us, for any positive integer $n$, 
to write
\begin{equation}
\Delta ^n(T_c)= \lim_{\eta \rightarrow 0} \sum _{j=1}^n
a_{nj} \, \eta ^{j+2-\frac {j+1}{2}D},
\label{end}
\end{equation}
which completes the proof.
\end{appendix}


\begin{figure}
\caption{ Heat capacity measurements for helium-4  in various porous
media.
(a) Jewler's rouge (powder) [1],
(b) Vycor glass [2],
(c) Grafoil  [3].
(d) Xerogel  [4].  Some data have been deleted for clarity. 
  Arrows indicate temperatures below which the film 
relaxation times go to zero.} 
\end{figure}
\begin{figure}
\caption{ Superfluid density of Helium-4  in several porous media.
The sample E (0.133 $g/cm^3$ DESY aerogel) and F (0.200 $g/cm^3$ 
Air glass aerogel) are aerogels of two different porous structures.
 The solid lines represent superfluid fraction in bulk helium. [8,6] }
\end{figure}
\begin{figure}
\caption{ The critical temperatures of the ideal Bose gas system
 between $D=2$ and $D=3$.  
The slop  is almost linear as observed in the experiments. }
\end{figure}
\begin{figure}
\caption{ Specific heat functions of the ideal Bose gas system between
 2 and 3 dimensions.  The unit of temperature is $T_o$ where
 $T_o \simeq 5.42 K   \left[ = g_{D/2}(1) ^{\frac{2}{D}} T_c(D) \right] $.  }
 \end{figure}
\begin{figure}
\caption{ Plot of the first derivative of the specific heat
 functions of the ideal Bose gas between $D=2$ and $D=3$. 
There are no cusps when $D < 3$. The unit of y-axis is arbitrary.  }
\end{figure}
\begin{figure}
\caption{ The condensate fraction in $D$-dimensional space.
$D=2.6, \;2.7,\; 2.8,\; 2.9,\; 3.0$ from left to right.}
\end{figure}
\mediumtext
\begin{table}
\caption{The hierarchy of the superfluidity transition between $D=2$ and
$D=3$.  The symbol {\em `c'} stands for being
 continuous at $T_c$, and  {\em `d'} for being
 discontinuous at $T_c$. The {\em Class} stands 
for the class of function.}
$$
\vbox{\tabskip 1em plus 2em minus .5em
\halign to 400pt{\hfil #\hfil && #\hfil \cr
\noalign{\hrule} \noalign{\hrule}
{}~ &~ &~ \cr
Dimension  & $C_V$ &
$\big( \frac{\partial}{\partial T} \big)   C_V$  &
$\big( \frac{\partial}{\partial T} \big)^2 C_V$  &
$\big( \frac{\partial}{\partial T} \big)^3 C_V$  &
$\big( \frac{\partial}{\partial T} \big)^4 C_V$  &
$\cdots$ & \em {Class} \cr 
\noalign{\hrule}
{}~ &~ &~ \cr
$  D = 3 $ &
\em c & \em d  & & & & &  $ \em {C^0} $ \cr
$\frac {8}{3} \leq D < \frac {6}{2} $ &
\em c & \em c & \em d   & & & &  $ \em {C^1} $ \cr
$\frac {10}{4} \leq D < \frac {8}{3} $ &
\em c & \em c & \em c & \em d   & & &  $ \em {C^2} $ \cr
$\vdots$ & & & & & & & $\vdots$ \cr
$\frac {2(j+2)}{j+1} \leq D < \frac {2(j+1)}{j} $ &
\em c & \em c & \em c & \em c &  $\cdots$ \,\, 
({\em d}) & &  $ \em {C^{j-1}} $ \cr
$\vdots$ & & & & & & & $\vdots$ \cr
$ D = 2 $ &
\em c & \em c & \em c & \em c & \em c   & $\cdots$ &  $\em {C^\infty}$ \cr
{}~ &~ &~ \cr
\noalign{\hrule} \noalign{\hrule}
}} $$
\end{table}

\begin{references}
\bibitem{Fred} H.P.R. Frederikse, Physica, {\bf 15}, 860 (1949).
\bibitem{Brew} D.F. Brewer, J. Low Temp. Phys.  {\bf 3}, 205  (1970).
\bibitem{Bret} M. Bretz, Phys. Rev. Lett.  {\bf 31}, 1447 (1973).
\bibitem{Fino} D. Finotello, K.A. Gillis, A. Wong, and M.H.W. Chan,
	       Phys. Rev. Lett.  {\bf 61} (1988).
\bibitem{Steele} L.M. Steele and D. Finotello, J. Low Temp. Phys.
		  {\bf 89}, 645  (1992). 
\bibitem{Rep1} J.D. Reppy,  {\it Phase Transitions in Surface Films},
	       edited by J.G. Dash and J. Ruvalds  (Plenum, New York, 
	       1980) p. 233.
\bibitem{Rep2} J.D. Reppy, J. Low Temp. Phys.  {\bf 87}, 205  (1992).
\bibitem{Rep3} G.K.S. Wong, P.A. Crowell, H.A. Cho, and J. D. Reppy,
	 Phys. Rev. B. {\bf 48}, 3858 (1993).
\bibitem{Bowl} R.H. Bowley and S. Giorgini,
	 J. Low Temp. Phys.  {\bf 93}, 987 (1993).
\bibitem{Khur} A. Khurana, Physics Today,  {\bf 42}, 21 (1989).
\bibitem{Gob} D.F. Goble and L.E.H. Trainor, Can. J. Phys.
	       {\bf 44}, 27  (1966).
\bibitem{KT} J.M. Kosterlitz and D.J. Thouless, J. Phys. C
		  {\bf 6}, 1181  (1973). 
\bibitem{Rep4} P.A. Crowell, F.W. Van Keuls, J.D. Reppy,
	 Phys. Rev. B. {\bf 55}, 12620 (1997).
\bibitem{phil} J.A. Phillips, D. Ross, P. Taborek, and J.E. Rutledge,	
	Phys. Rev. B. {\bf 58}, 3361 (1998).
\bibitem{Gasp} F.M. Gaspirini and S. Mhlanga, Phys. Rev. B.  {\bf 33}, 
	       5066 (1986).
\bibitem{Will} G.A. Williams, Phys. Rev. Lett.  {\bf 64}, 978  (1990).
\bibitem{Saar} M. Saarela, B.E. Clements, E. Krotscheck, 
	and F.V. Kusmartsev, J. Low Temp. Phys.  {\bf 93}, 971 (1993).
\bibitem{PfOb} P. Pfeifer and M. Obert,   {\it The Fractal Approach
	       to Heterogeneous Chemistry}, edited by D. Avnir
	       (Wiely, Chichester, 1989) p. 11. 
\bibitem{PfWu} P. Pfeifer, Y.J. Wu, M.W. Cole, and J. Krim, Phys. Rev. Lett. 
	       {\bf 62}, 1977  (1989).
\bibitem{PfCo} P. Pfeifer and M.W. Cole, New J. Chem.  {\bf 14}, 221  (1990).
\bibitem{Pf} P. Pfeifer,  {\it Chemistry and Physics of Solid Surfaces},
		edited by R. Vanselow and R. Howe
	       (Springer-Verlag, Berlin, 1988) Vol. VII, p. 283.
\bibitem{Ziff} R.M. Ziff, G.E. Uhlenbeck, and M. Kac, Phys. Rep. {\bf 32},
	169 (North-Holland, Amsterdam, 1977).
 \bibitem{kim} S.-H. Kim, Int. J. Bif. Cha. {\bf 7}, 1053 (1997).
 \bibitem{Robi} J.E. Robinson, Phys. Rev.  {\bf 83}, 678  (1951).
\bibitem{InT}   {\it International Critical Tables of Numerical Data}
	       (McGraw-Hill, New York, 1933) I-102.
\bibitem{Grad} I.S. Gradshteyn and I.M. Ryzhik,  {\it Tables of Integrals,
	     series and products}  (Academic, New York, 1980) p. 946.
\end{references}
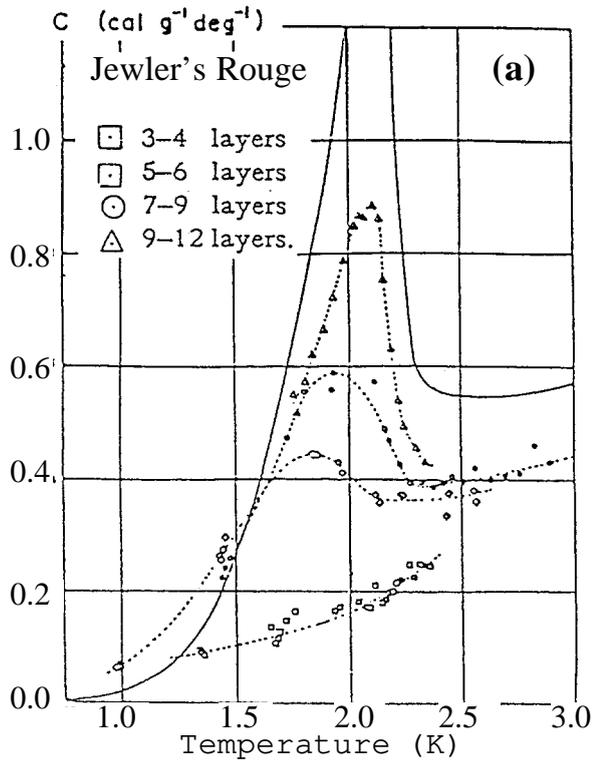
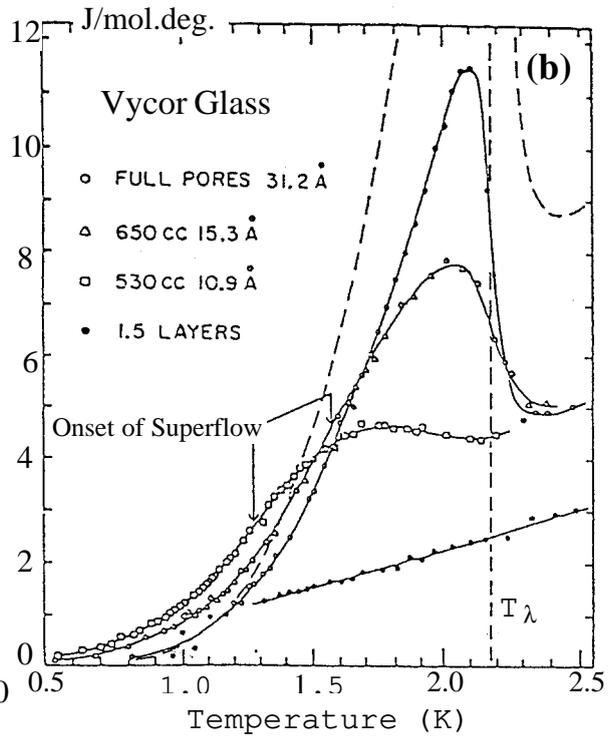
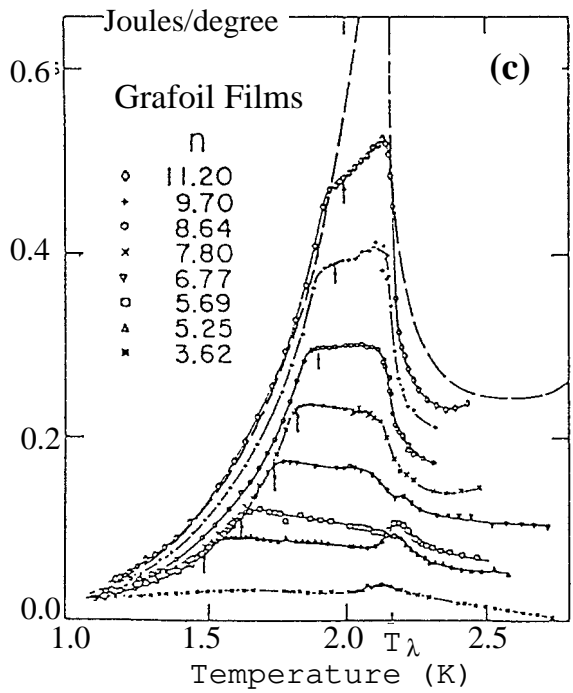
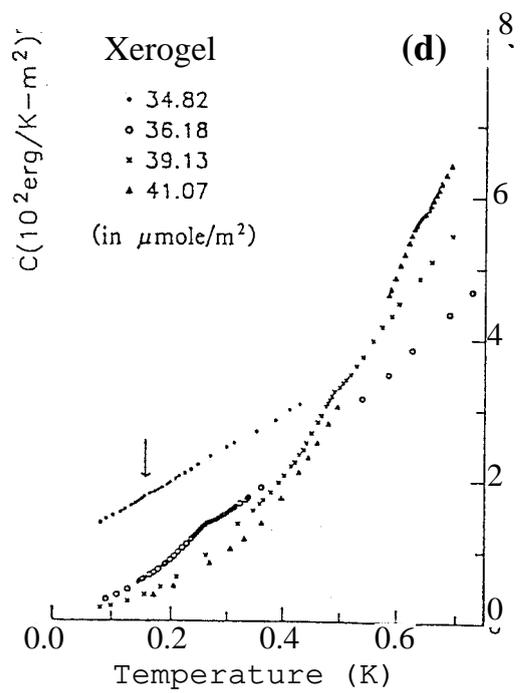
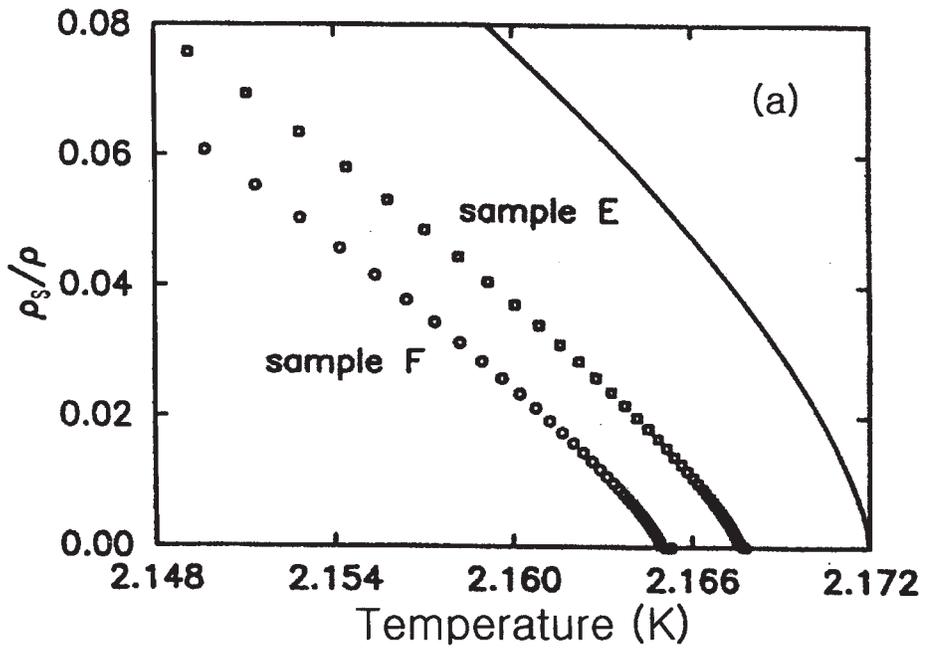
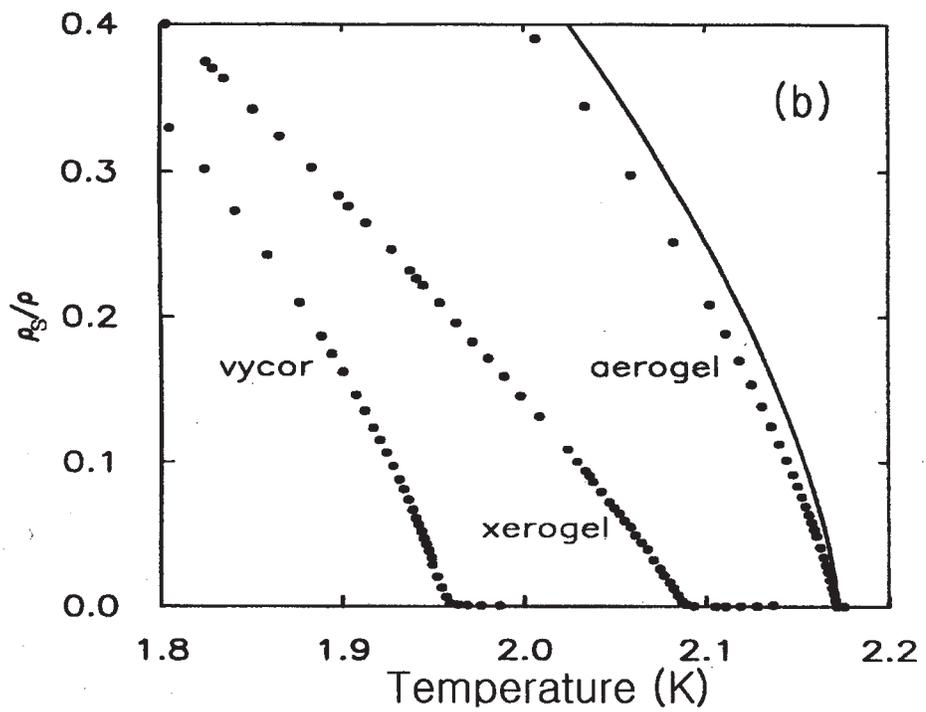
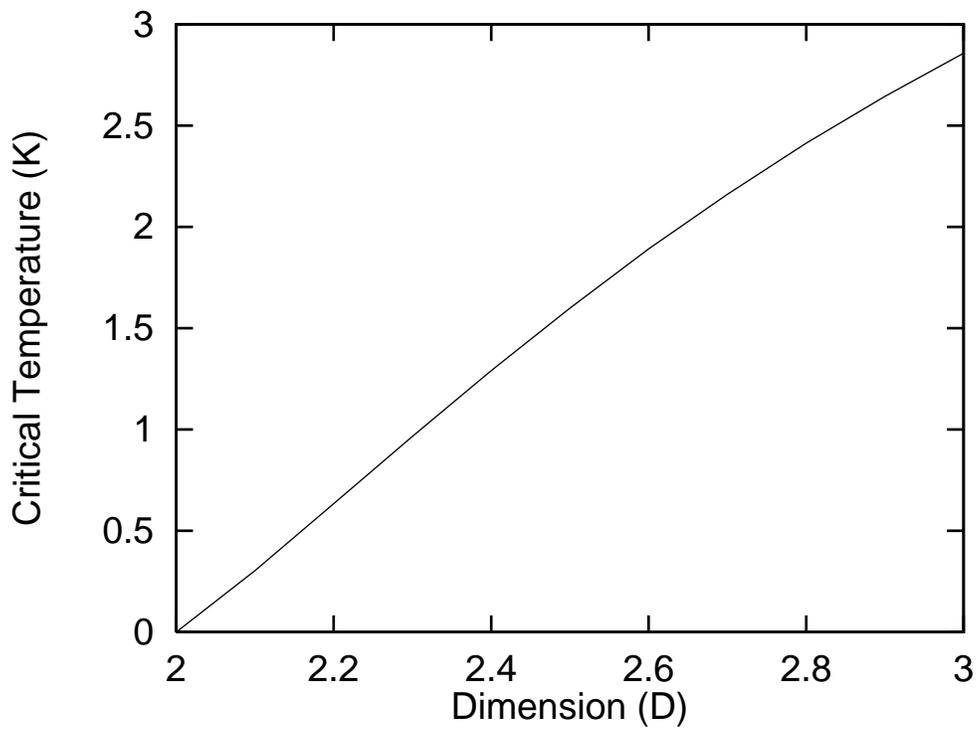
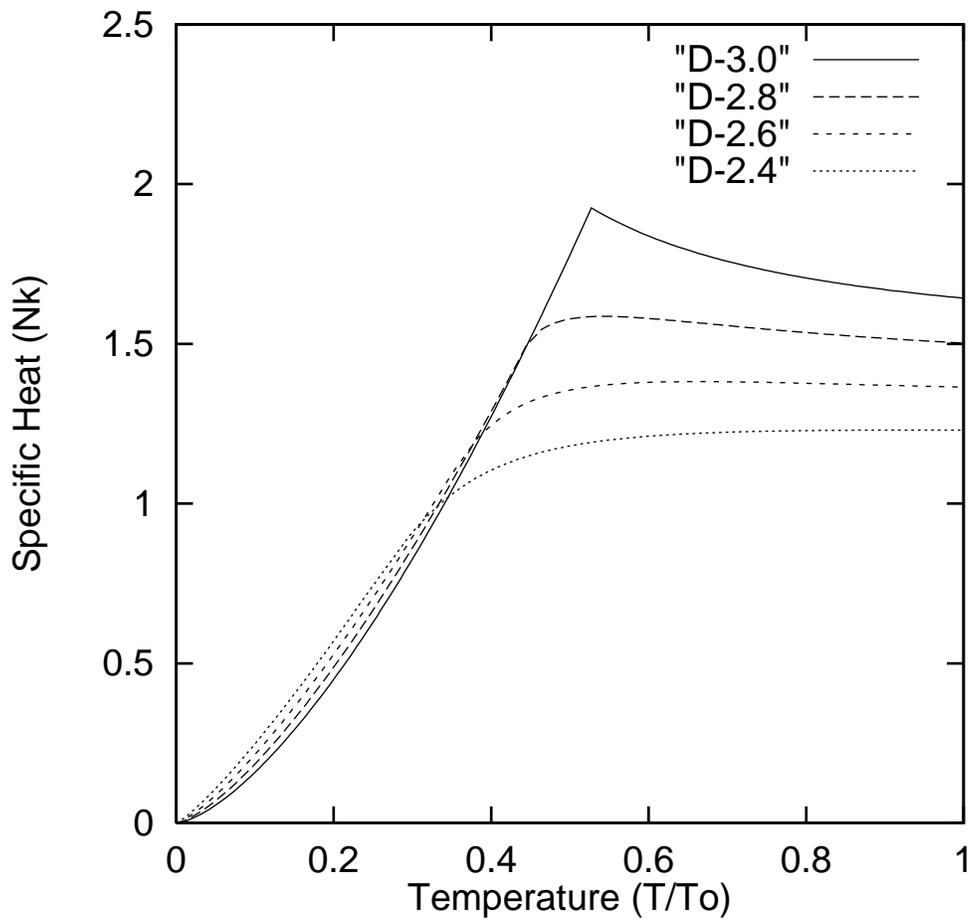
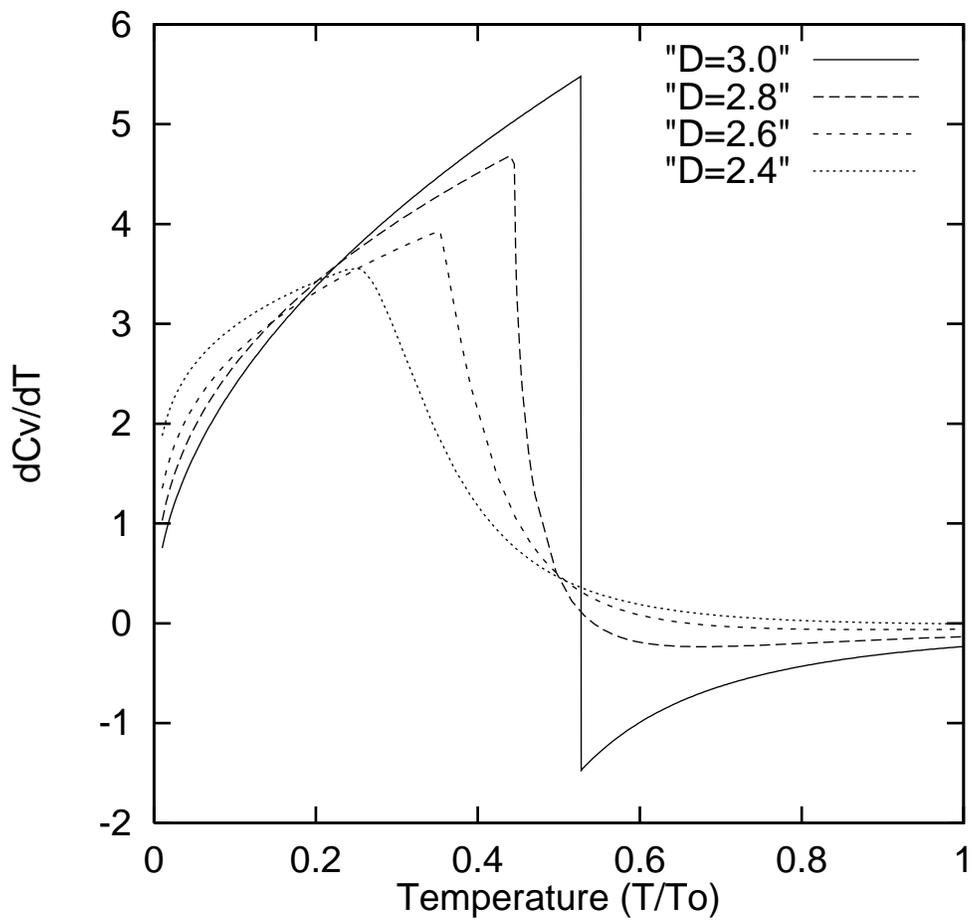
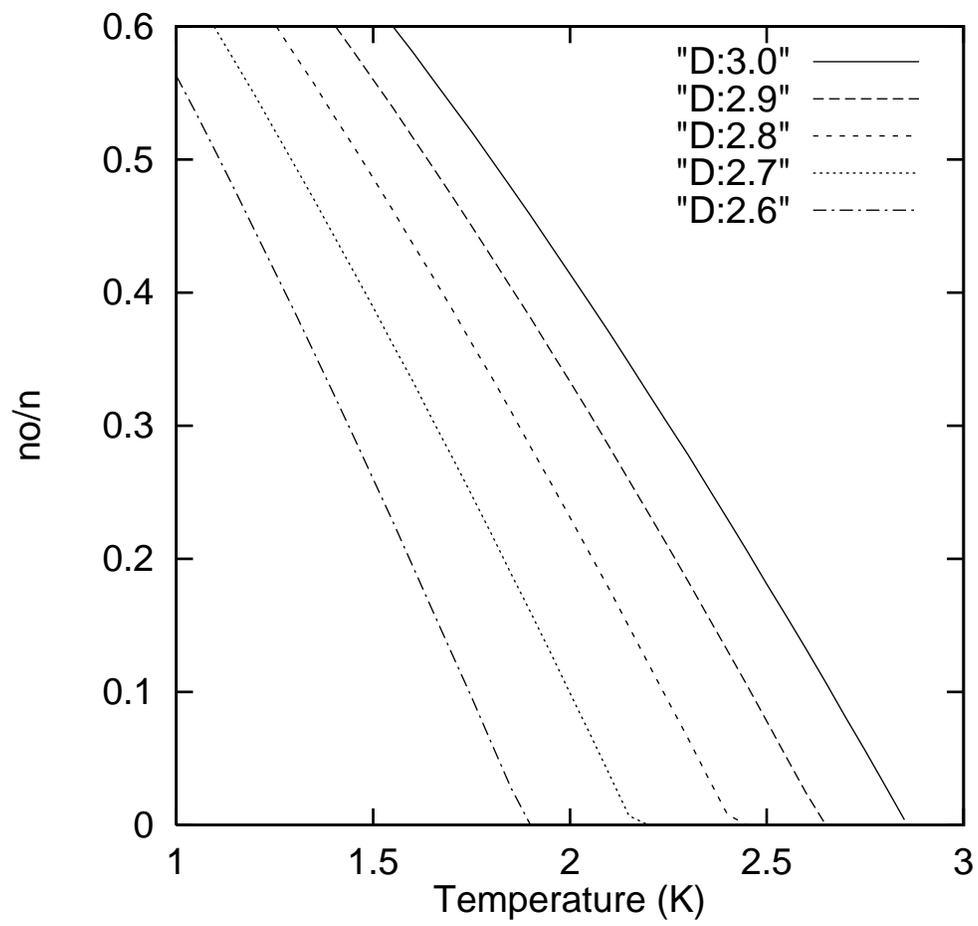
\end{document}